\documentclass[sigconf]{acmart}
\AtBeginDocument{%
  }

\usepackage{amsthm,amsmath,bm}
\usepackage{multirow}
\usepackage{mathrsfs}
\usepackage{graphicx}
\usepackage{subfigure} 
\usepackage{float}
\usepackage[export]{adjustbox} 
\usepackage{xspace}
\graphicspath{ {./Figs/} }

\usepackage{algorithm}
\usepackage{algorithmic}
\usepackage{balance}
\usepackage{enumitem}

\newcommand{\eat}[1]{}
\newcommand{\paratitle}[1]{\vspace{1.5ex}\noindent\textbf{#1}}

\newcommand{\baby}{LMN\xspace}

\newcommand{\ignore}[1]{}

\copyrightyear{2025}
\acmYear{2025}
\setcopyright{acmlicensed}\acmConference[WWW Companion '25]{Companion Proceedings of the ACM Web Conference 2025}{April 28-May 2, 2025}{Sydney, NSW, Australia}
\acmBooktitle{Companion Proceedings of the ACM Web Conference 2025 (WWW Companion '25), April 28-May 2, 2025, Sydney, NSW, Australia}
\acmDOI{10.1145/3701716.3715514}
\acmISBN{979-8-4007-1331-6/2025/04}

\begin{document}

\title{Large Memory Network for Recommendation}

\author{Hui Lu}
\email{luhui.xx@bytedance.com}
\affiliation{%
  \institution{ByteDance}
  \city{Hangzhou}
  \country{China}
}

\author{Zheng Chai\textsuperscript{†}}
\thanks{\textsuperscript{†}Corresponding Author.}
\email{chaizheng.cz@bytedance.com}
\affiliation{%
  \institution{ByteDance}
  \city{Hangzhou}
  \country{China}
}

\author{Yuchao Zheng}
\email{zhengyuchao.yc@bytedance.com}
\affiliation{%
  \institution{ByteDance}
  \city{Beijing}
  \country{China}
}

\author{Zhe Chen}
\email{chenzhe.john@bytedance.com}
\affiliation{%
  \institution{ByteDance}
  \city{Beijing}
  \country{China}
}

\author{Deping Xie}
\email{deping.xie@bytedance.com}
\affiliation{%
  \institution{ByteDance}
  \city{San Jose}
  \country{USA}
}

\author{Peng Xu}
\email{xupeng@bytedance.com}
\affiliation{%
  \institution{ByteDance}
  \city{San Jose}
  \country{USA}
}

\author{Xun Zhou}
\email{zhouxun@bytedance.com}
\affiliation{%
  \institution{ByteDance}
  \city{Beijing}
  \country{China}
}

\author{Di Wu}
\email{di.wu@bytedance.com}
\affiliation{%
  \institution{ByteDance}
  \city{Beijing}
  \country{China}
}

\renewcommand{\shortauthors}{Hui Lu et al.}

\begin{abstract}
    Modeling user behavior sequences in recommender systems is essential for understanding user preferences over time, enabling personalized and accurate recommendations for improving user retention and enhancing business values. Despite its significance, there are two challenges for current sequential modeling approaches. From the spatial dimension, it is difficult to mutually perceive similar users' interests for a generalized intention understanding; from the temporal dimension, current methods are generally prone to forgetting long-term interests due to the fixed-length input sequence. In this paper, we present Large Memory Network (LMN), providing a novel idea by compressing and storing user history behavior information in a large-scale memory block. With the elaborated online deployment strategy, the memory block can be easily scaled up to million-scale in the industry. Extensive offline comparison experiments, memory scaling up experiments, and online A/B test on Douyin E-Commerce Search (ECS) are performed, validating the superior performance of LMN. Currently, LMN has been fully deployed in Douyin ECS, serving millions of users each day.
\end{abstract}

\ccsdesc[500]{Information systems~Recommender systems}

\keywords{Large Recommender Models, Memory Learning, Personalized Recommender System, Sequential Modeling}

\maketitle

\section{Introduction}
User behavior sequences are a crucial means of depicting user interests and preferences \cite{chai2022user, de2021transformers4rec}. Currently, many outstanding sequential modeling methods have been approached, which can be divided into two branches. The first one is to adopt a target-attention mechanism for short sequence modeling, for example, DIN \cite{zhou2018deep}, DIEN \cite{zhou2019deep}, CAN \cite{bian2022can}. As such methods are mostly developed for short sequences, the other branch develops search-based mechanism to perceive longer sequences in the model, like SIM \cite{pi2020search}, TWIN \cite{chang2023twin}, etc. However, much efforts are generally needed to deploy such complex two-stage models for online serving, and the behavior discarding further results in incomplete interest estimation. Therefore, another type of approach that is emerging as a hotspot is to directly model long sequence in the recommendation model \cite{liu2023deep, cao2022sdim, yu2024ifa, zhaiactions}, which inevitably increases the computational cost of model training and inference, introducing a substantial online overhead for industrial recommenders.

Currently, scaling laws have guided the language model design by continuously scaling up the model parameters, showing promising performance gains. The discovery has also spurred research in other fields towards increasing model complexity, for example, increasing the dense network parameters \cite{zhangwukong} and sparse embedding parameters \cite{guoembedding} in industrial recommenders. However, in practice, simply scaling model parameters can directly lead to decreased training and inference efficiency, and the computational FLOPs increment further limits the model scalability. To mitigate the computational and GPU memory overhead, the memory network structure can amplify model parameters through memory mechanism to enhance model capacity, which has been applied in the field of natural language modeling. The Key-Value Memory Network \cite{miller2016key} proposed by Miller et al. abstracts the basic structure of memory network mechanism, which mainly consists of query, keys, and values, representing an early form of attention mechanism. However, with the increase of memory space, both the computational complexity and memory access overhead of such basic key-value memory structure will also increase. Under hardware constraints, the memory space cannot be infinitely enlarged, limiting the expressive power and scaling laws of memory. To this end, Lample et al. \cite{lample2019large} developed large memory layers by decomposing keys to decrease the complexity. Despite the popularity and effectiveness, such memory mechanism is rarely used in the field of recommendation. Pi et al. proposed a multi-channel user interest memory network (MIMN) by introducing neural Turing machine and memory induction unit \cite{pi2019practice}, while the complex serial modeling structure with GRU made the model training inefficient, and the limited memory space further restricts its effectiveness.

To solve the above problems, in this paper, we present Large Memory Network (LMN), a lightweight yet effective sequential modeling framework for user intention understanding. Generally, the overall contributions can be summarized as follows:
\begin{itemize}
    \item We present a novel lightweight large memory network, by which the user sequences can be effectively compressed into memory parameters and achieve large-scale expansion of model parameter capacity with little computational cost.
    \item The designed LMN builds a large global-shared table for different users, enabling spatial perception among users and temporal memory of long-term user interests. Both a product quantization-based memory decomposition and an industrial online framework are provided for efficient deployment.
    \item Extensive comparison and scaling up experiments, and online A/B experiment illustrate the superiority. LMN has been fully deployed in Douyin E-Commerce, serving millions of users.
\end{itemize}

\section{Methodology}
\subsection{Preliminaries}
\subsubsection{Problem Formulation}
This paper focuses on the click-through rate (CTR) prediction tasks in recommendation, and the overall learning objective can be formulated as follows:

\begin{equation}
    \hat{y} = f(\bm{u}, \bm{i}, \bm{c}, \bm{s})
\end{equation}
where $\bm{u}$, $\bm{i}$, $\bm{c}$ denote user-side, item-side, and user-item cross feature embeddings, respectively, and $\bm{s}$ represents the user sequence representation. $f$ denotes the recommendation model for predicting the CTR $\hat{y}$.

\subsection{The Proposed \baby}
\begin{figure}[h]
\center
\includegraphics[width=\linewidth]{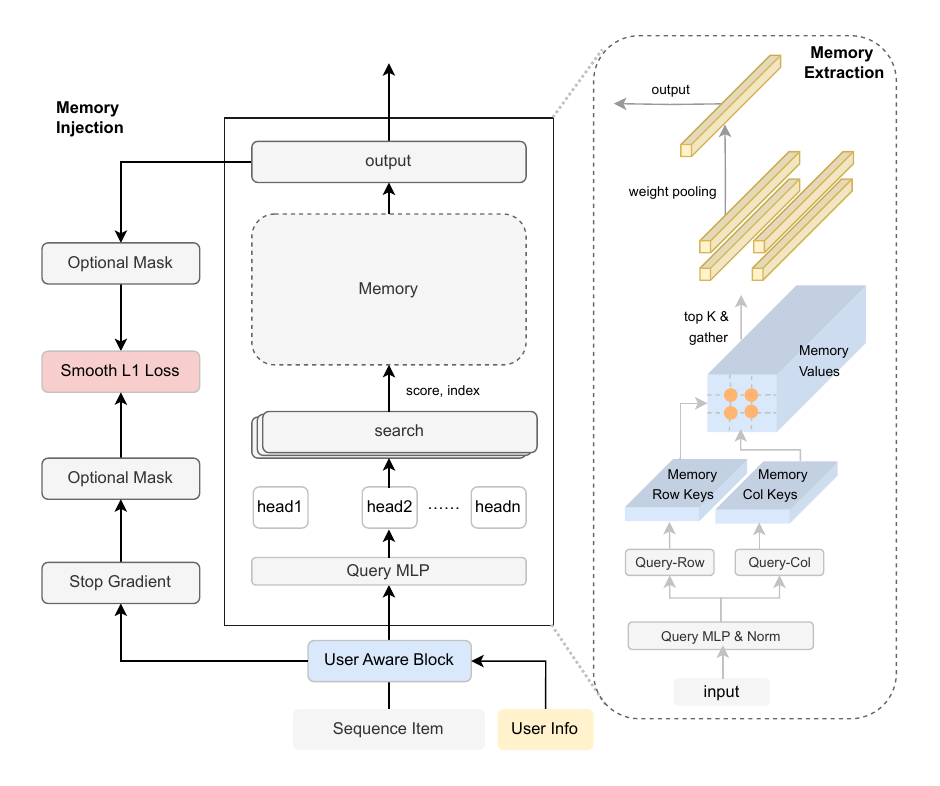}
\caption{Overall framework of the proposed \baby. There are two learnable components in the memory block including memory keys and memory values. The \baby uses a two-axis activation for locating the top $K$ relevant values for a given specific query. Besides, the Smooth L1 loss is performed to inject historical interaction with corresponding user information into the memory value slots. The optional mask is used as some of the sequence items are padded to a fixed length with zero embeddings in practice.}
\label{fig:overview}
\end{figure}

The structure of the proposed \baby is illustrated in Figure~\ref{fig:overview}. In general, LMN is an end-to-end storage unit that compresses and memorizes historical behavior sequences and user information.

\subsubsection{Overall Framework}
The learning objective of \baby is to build a large-scale memory block and find the top-$K$ most relevant memory slots in the block for assisting user intention modeling. It is noted that all users share the same memory block, allowing user behavior information to generalize in a much larger parameter space (memory), which establishes perception among users spatially and memorizes long-term user interest temporally. As shown in Figure~\ref{fig:overview}, the overall framework of the LMN has two stages, i.e., memory extraction and memory injection, which will be introduced in detail.

\subsubsection{Memory Extraction}
Generally, denote the memory key as $M_K \in \mathbb{R}^{n \times d}$, and the memory value as $M_V \in \mathbb{R}^{n \times d}$, where $n$ and $d$ represent the number and dimension of the memory slots. Given an interacted item $x \in \mathbb{R}^{1 \times d}$ in user sequence as the memory query, the calculation process of memory activation can be written as:

\begin{align}
M_O = Softmax (x \cdot M_K^{\mathrm{T}}) \cdot M_V
\label{eqn:overall}
\end{align}

\noindent where $M_O$ denotes the output of the memory block. 

As shown in Eq.~\ref{eqn:overall}, the computational complexity of the similarity score, i.e., $s = x \cdot M_K^{\mathrm{T}}$, is $\bm{\mathcal{O}}(nd)$ which is linearly proportional to the scale of the number of memory slots $n$. To further decrease the computational cost, the memory activation mechanism is designed based on product quantization \cite{jegou2010product, lample2019large}. Specifically, we quantize the original $M_K \in \mathbb{R}^{n \times d}$ as:

\begin{align}
{M_K} = {M_{K-row}} \times {M_{K-col}}
\label{eqn:quantize}
\end{align}

\noindent where we decompose the original $M_K$ with $n$ memory slots into two subgroups, i.e., $M_{K-row} \in \mathbb{R}^{\sqrt{n} \times d}$ and $M_{K-col} \in \mathbb{R}^{\sqrt{n} \times d}$, corresponding to the row index keys and column index keys, respectively. "$\times$" here denotes a cross operation. With both the row and col indices, the activated memory values in $n$ memory value slots can be located, as illustrated in Fig. 1. As each subgroup includes $\sqrt{n}$ memory slots, the computational cost of the attention score calculation in Eq.~\ref{eqn:overall} can be decreased from $\bm{\mathcal{O}}(nd)$ to $\bm{\mathcal{O}}(\sqrt{n}d)$ effectively.

\textbf{User-Aware Block.} Correspondingly, to lookup a memory value for a given interacted-item $x$, we first obtain the corresponding queries $M_{Q-row}$ and $M_{Q-col}$ for both keys in Eq.~\ref{eqn:quantize}. However, due to the highly personalized characteristics of users in recommenders, the same representation of an item $x$ exhibits varying semantics across various users. Thus, relying only on the similarity of item embedding as a metric is insufficient. To this end, we present user-aware block, which introduces user-aware information into the queries for better generalization performance:

\begin{align}
M_Q = MLP(Merge(x, \bm{u}))
\label{eqn:user-aware}
\end{align}

\noindent where $\bm{u}$ is the user-side features like the user id or demographic features, and the Merge operation means the add or concatenation operation. In this paper, concatenation is used. Then, the two queries can be generated corresponding to the row keys and column keys:

\begin{align}
{M_{Q-row}} = MLP(M_Q)\\
{M_{Q-col}} = MLP(M_Q)
\label{eqn:query-mlp}
\end{align}

Based on the $M_{Q-row} \in \mathbb{R}^{1 \times d}$ and $M_{Q-col} \in \mathbb{R}^{1 \times d}$, we obtain the memory activation scores at the row axis, column axis, and the overall:

\begin{align}
S_{row} &= M_{Q-row} \cdot M_{K-row}^{\mathrm{T}}\\
S_{col} &= M_{Q-col} \cdot M_{K-col}^{\mathrm{T}}\\
S &= Broadcast(S_{row}, S_{col})
\label{eqn:s}
\end{align}

\noindent where both $S_{row}$ and $S_{col} \in \mathbb{R}^{1 \times \sqrt{n}}$, and the broadcast operation means that $S_{row}$ and $S_{col}$ are first reshaped as $S_{row} \in \mathbb{R}^{1 \times \sqrt{n}}$ and $S_{col} \in \mathbb{R}^{\sqrt{n} \times 1}$, then the two scores are added with the broadcast mechanism\footnote{https://www.tensorflow.org/guide/tensor\#broadcasting} to obtain a final score at $S \in \mathbb{R}^{\sqrt{n} \times \sqrt{n}}$, which is finally reshaped to $\mathbb{R}^{n}$.

With the memory activation score $S$, we select the top $K$ highest scores to lookup the values and gather them with a weighted-sum mechanism:

\begin{align}
M_O = Softmax(S_{TopK}) \cdot M_{V-Top K}
\label{eqn:weighted-sum}
\end{align}
in which $S_{Top K}$ denotes the highest $K$ activation scores in $S$ and $M_{V-Top K}$ represents the top $K$ score indexed-values.

To further increase the model capacity and improve performance, both the query and keys can be obtained via a multi-head mechanism.

\subsubsection{Memory Injection}The stage of injection aims to memorize compressed information into the value table. Traditional memory structures might use a write-back operation to inject information \cite{pi2019practice}. However, under the query-key activation framework, a more reasonable solution is to use the gradient-descent approach to update both the key and value memory tables. Furthermore, to memorize the information of the user-item pairs in the designed memory slots, a Smooth-L1 \cite{girshick2015fast} based memory loss is adopted to inject user-item pair information into the memory slots:

\begin{align}
Loss_{Memory} = SmoothL1(M_O, M_Q)
\label{eqn:mem-loss}
\end{align}

\subsubsection{Model training and deployment} The LMN is a lightweight module while with the potential to significantly increase the model parameter capacity. It can be used as a plug-and-play module for current CTR prediction models, and the overall loss can be obtained as follows with a balanced parameter $\alpha$:

\begin{align}
Loss_{LMN} = Loss_{CTR} + \alpha Loss_{Memory}
\label{eqn:loss}
\end{align}
\noindent where $Loss_{CTR}$ is the loss value of the original CTR task.

\begin{figure}[h]
\center
\includegraphics[width=\linewidth, trim=0 20 0 17,clip]{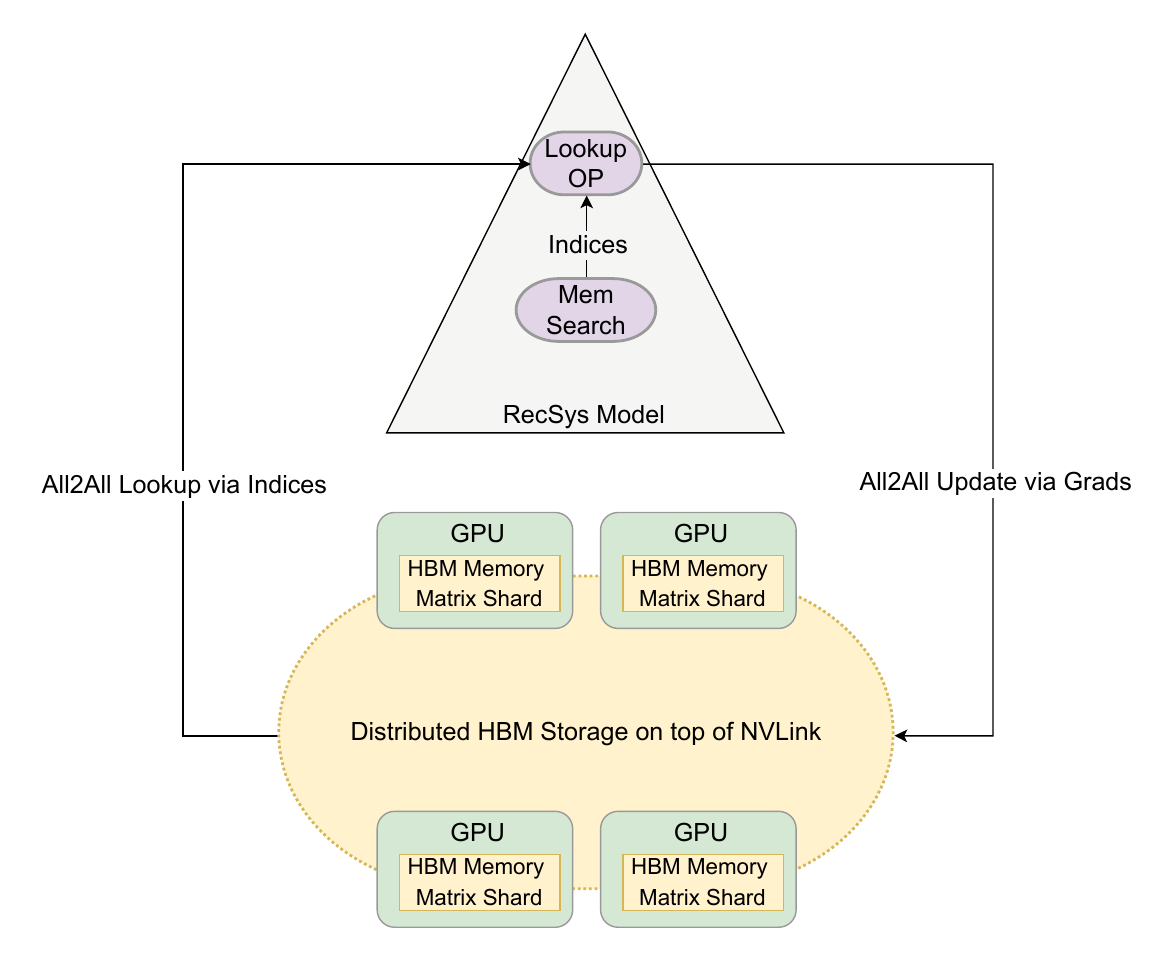}
\caption{Online deployment framework of memory parameter server (MPS) for the proposed \baby at ByteDance.}
\label{fig:sys}
\end{figure}

To deploy the LMN in online scenarios, it is noted that among the two modules including the memory keys and memory values, the latter one has a much larger capacity which should be carefully designed for deployment in industrial applications. Specifically, we introduce memory parameter server (MPS), which adopts a GPU-sharding strategy for distributed storage of the memory values on high bandwidth memory (HBM) of multi-GPUs. The detailed framework of MPS is illustrated in Fig.~\ref{fig:sys}. In the model training process, the recommendation model launches a request for lookup of corresponding values according to the memory activation score $S$. Then, the all2all lookup is performed for distributed HBM storage on top of GPU NVLink\footnote{https://www.nvidia.com/en-us/data-center/nvlink/}. Then, the memory values are updated with the all2all gradient updates. For serving, only the All2All lookup is reserved for memory search.

\section{Experiment}
\subsection{Evaluation Setting}
\paratitle{Dataset}. To investigate the performance of \baby, we conduct experiment at Douyin E-Commerce Search (ECS), which serves as the most important scenario to meet the e-commerce demands of users in search scenarios at ByteDance and attracts millions of page views of online traffic each day. We collect and sample online traffic logs from Oct. 1st, 2024 to Oct. 28th, 2024, with 4 weeks and \textbf{3.6 billion} samples, and the former 3 weeks are used for training and the last week is for evaluation. 

\paratitle{Comparison Methods and Metrics}. We select the following SOTA methods for evaluating the effectiveness of the proposed LMN. The \textbf{DNN} baseline is a typical DNN-based structure for CTR prediction. The basic sequential modeling methods include \textbf{Pooling} and \textbf{DIN} \cite{zhou2018deep} conducted on user sequence with click behaviors whose length and embedding size are 50 and 32, respectively. Based on DIN, we further introduce \textbf{SIM} \cite{pi2020search}, \textbf{MIMN} \cite{pi2019practice}, and the proposed \textbf{\baby}. The number of memory keys is set to 300. Both the AUC and LogLoss metrics are used to evaluate the ranking model performance. Further, we introduce relative improvement (Imp.) to measure the relative AUC gain \cite{yan2014coupled}.

\subsection{Experimental Results}
\paratitle{Comparison Results}. The performance of different methods is summarized in Table~\ref{tab:auc}. There are the following observations. First, it is observed that sequence modeling is of great significance in recommendation, and the basic Pooling method achieves 1.32\% AUC improvement compared with the DNN baseline. Then, the personalized weights in DIN are more effective than Pooling, and it is observed that modeling a much longer sequence using SIM achieves better performance than DIN (0.7348 vs. 0.7327). Finally, compared to MIMN, the proposed \baby shows prominent effectiveness (4.68\% and 1.46\% improvements in AUC and LogLoss). We believe that the reason is due to the effectiveness of the user-aware memory learning mechanism that enables both spatial perception and temporal memorization. 

\paratitle{Performance of Scaling Up Memories}. Further, to validate the performance when scaling up the memory, we vary the number of memory keys from [50, 100, 200, 300, 500], and the results are illustrated in Table~\ref{tab:scalingup}. From the table, it can be observed that both performance metrics consistently improve with the number of keys scales up.

\begin{table}[]
\caption{Comparison results of different methods on Douyin E-Commerce Search Dataset. Boldface denotes the best results with a confidence level at $p < 0.05$.}
\begin{tabular}{ccccc}
\hline
        & AUC $\uparrow$ & AUC Imp. & LogLoss $\downarrow$ & LogLoss Imp. \\ \hline
Base    & 0.7265 & -        & 0.6075  & -            \\
Pooling & 0.7295 & +1.32\%  & 0.6053  & -0.36\%      \\
DIN     & 0.7327 & +2.34\%  & 0.6023  & -0.86\%      \\
SIM     & 0.7348 & +3.66\%  & 0.6009  & -1.09\%      \\
MIMN    & 0.7363 & +4.33\%  & 0.5991  & -1.39\%      \\
\textbf{\baby}     & \textbf{0.7371} & \textbf{+4.68\%}  & \textbf{0.5986}  & \textbf{-1.46\%}      \\ \hline
\end{tabular}
\label{tab:auc}
\end{table}

\begin{table}[]
\caption{Model performance with the number of memory keys scaling up.}
\begin{tabular}{ccccc}
\hline
        & AUC $\uparrow$ & AUC Imp. & LogLoss $\downarrow$ & LogLoss Imp. \\ \hline
Base    & 0.7265 & -        & 0.6075  & -            \\
\baby ($\sqrt{n}$=50) & 0.7357 & +4.06\%  & 0.5997  & -1.28\%      \\
\baby ($\sqrt{n}$=100)     & 0.7364 & +4.37\%  & 0.5993  & -1.35\%      \\
\baby ($\sqrt{n}$=200)     & 0.7370 & +4.64\%  & 0.5987  & -1.45\%      \\
\baby ($\sqrt{n}$=300)    & 0.7371 & +4.68\%  & 0.5986  & -1.46\%      \\
\baby ($\sqrt{n}$=500)     & 0.7380 & +5.08\%  & 0.5978  & -1.60\%      \\ \hline
\end{tabular}
\label{tab:scalingup}
\end{table}

\subsection{Online A/B Experiment}
We perform online A/B test from 30th Aug. to 5th Sep. 2024, with 20\% online traffic on Douyin E-Commerce Search to validate the effectiveness, which hits more than 160 million users. The results are illustrated in Table~\ref{tab:ab}. It can be observed that after deploying the proposed \baby, the most influential commercial metrics, i.e., order/user and order/search, are raised by 0.87\% and 0.72\%, respectively, with a substantial significance of $p$=0. Besides, after deploying the \baby, there is only a slight increment of online serving latency at 0.38\%, showing the efficiency of the deployment framework.

\begin{table}[]
\caption{Online A/B test result of the proposed \baby on Douyin E-Commerce Search. The boldface denotes that the improvement is significant with $p = $\textbf{0}.}
\begin{tabular}{ccc}
\hline
         & Order/User & Order/Search \\ \hline
Baseline & +0.00\%          & +0.00\%            \\
\textbf{\baby}      & \textbf{+0.87\%}    & \textbf{+0.72\%}      \\ \hline
\end{tabular}
\label{tab:ab}
\end{table}

\section{Conclusion}
In this paper, we present Large Memory Network (\baby), a novel memory-enhanced user sequential modeling and intention understanding framework for recommendation. It compresses and memorizes user interest through a large-scale memory block. With such design, the memory can 1) spatially perceive different users' sequences for better generalization, and 2) temporally memorize user long-term interest through a user-aware block. Besides, to further reduce computational cost and deployment overhead, both a product quantization-based memory decomposition and an industrial online memory parameter server framework are devised for industrial online deployment. Currently, \baby has been fully deployed at ByteDance, which serves the major online traffic of E-Commerce Search service at Douyin.

It is noted that instead of memorizing the user's historical behavioral items in the current work, it is a promising direction to memorize any important information within the industrial recommender models, for example, memorizing the output of the feature-cross modules, or the output of the high-level compressor MLPs, to achieve the goal of perceiving similar user's instances and memorizing the same user's instance and further improve the effectiveness of the recommendation model.

\balance
\bibliographystyle{ACM-Reference-Format}
\bibliography{ref}


\begin{thebibliography}{19}


\ifx \showCODEN    \undefined \def \showCODEN     #1{\unskip}     \fi
\ifx \showISBNx    \undefined \def \showISBNx     #1{\unskip}     \fi
\ifx \showISBNxiii \undefined \def \showISBNxiii  #1{\unskip}     \fi
\ifx \showISSN     \undefined \def \showISSN      #1{\unskip}     \fi
\ifx \showLCCN     \undefined \def \showLCCN      #1{\unskip}     \fi
\ifx \shownote     \undefined \def \shownote      #1{#1}          \fi
\ifx \showarticletitle \undefined \def \showarticletitle #1{#1}   \fi
\ifx \showURL      \undefined \def \showURL       {\relax}        \fi
\providecommand\bibfield[2]{#2}
\providecommand\bibinfo[2]{#2}
\providecommand\natexlab[1]{#1}
\providecommand\showeprint[2][]{arXiv:#2}

\bibitem[Bian et~al\mbox{.}(2022)]%
        {bian2022can}
\bibfield{author}{\bibinfo{person}{Weijie Bian}, \bibinfo{person}{Kailun Wu}, \bibinfo{person}{Lejian Ren}, \bibinfo{person}{Qi Pi}, \bibinfo{person}{Yujing Zhang}, \bibinfo{person}{Can Xiao}, \bibinfo{person}{Xiang-Rong Sheng}, \bibinfo{person}{Yong-Nan Zhu}, \bibinfo{person}{Zhangming Chan}, \bibinfo{person}{Na Mou}, {et~al\mbox{.}}} \bibinfo{year}{2022}\natexlab{}.
\newblock \showarticletitle{CAN: feature co-action network for click-through rate prediction}. In \bibinfo{booktitle}{\emph{Proceedings of the fifteenth ACM international conference on web search and data mining}}. \bibinfo{pages}{57--65}.
\newblock


\bibitem[Cao et~al\mbox{.}(2022)]%
        {cao2022sdim}
\bibfield{author}{\bibinfo{person}{Yue Cao}, \bibinfo{person}{Xiaojiang Zhou}, \bibinfo{person}{Jiaqi Feng}, \bibinfo{person}{Peihao Huang}, \bibinfo{person}{Yao Xiao}, \bibinfo{person}{Dayao Chen}, {and} \bibinfo{person}{Sheng Chen}.} \bibinfo{year}{2022}\natexlab{}.
\newblock \showarticletitle{Sampling is all you need on modeling long-term user behaviors for CTR prediction}. In \bibinfo{booktitle}{\emph{Proceedings of the 31st ACM International Conference on Information \& Knowledge Management}}. \bibinfo{pages}{2974--2983}.
\newblock


\bibitem[Chai et~al\mbox{.}(2022)]%
        {chai2022user}
\bibfield{author}{\bibinfo{person}{Zheng Chai}, \bibinfo{person}{Zhihong Chen}, \bibinfo{person}{Chenliang Li}, \bibinfo{person}{Rong Xiao}, \bibinfo{person}{Houyi Li}, \bibinfo{person}{Jiawei Wu}, \bibinfo{person}{Jingxu Chen}, {and} \bibinfo{person}{Haihong Tang}.} \bibinfo{year}{2022}\natexlab{}.
\newblock \showarticletitle{User-aware multi-interest learning for candidate matching in recommenders}. In \bibinfo{booktitle}{\emph{Proceedings of the 45th International ACM SIGIR Conference on Research and Development in Information Retrieval}}. \bibinfo{pages}{1326--1335}.
\newblock


\bibitem[Chang et~al\mbox{.}(2023)]%
        {chang2023twin}
\bibfield{author}{\bibinfo{person}{Jianxin Chang}, \bibinfo{person}{Chenbin Zhang}, \bibinfo{person}{Zhiyi Fu}, \bibinfo{person}{Xiaoxue Zang}, \bibinfo{person}{Lin Guan}, \bibinfo{person}{Jing Lu}, \bibinfo{person}{Yiqun Hui}, \bibinfo{person}{Dewei Leng}, \bibinfo{person}{Yanan Niu}, \bibinfo{person}{Yang Song}, {et~al\mbox{.}}} \bibinfo{year}{2023}\natexlab{}.
\newblock \showarticletitle{TWIN: TWo-stage interest network for lifelong user behavior modeling in CTR prediction at kuaishou}. In \bibinfo{booktitle}{\emph{Proceedings of the 29th ACM SIGKDD Conference on Knowledge Discovery and Data Mining}}. \bibinfo{pages}{3785--3794}.
\newblock


\bibitem[de~Souza Pereira~Moreira et~al\mbox{.}(2021)]%
        {de2021transformers4rec}
\bibfield{author}{\bibinfo{person}{Gabriel de Souza Pereira~Moreira}, \bibinfo{person}{Sara Rabhi}, \bibinfo{person}{Jeong~Min Lee}, \bibinfo{person}{Ronay Ak}, {and} \bibinfo{person}{Even Oldridge}.} \bibinfo{year}{2021}\natexlab{}.
\newblock \showarticletitle{Transformers4rec: Bridging the gap between nlp and sequential/session-based recommendation}. In \bibinfo{booktitle}{\emph{Proceedings of the 15th ACM conference on recommender systems}}. \bibinfo{pages}{143--153}.
\newblock


\bibitem[Girshick(2015)]%
        {girshick2015fast}
\bibfield{author}{\bibinfo{person}{R Girshick}.} \bibinfo{year}{2015}\natexlab{}.
\newblock \showarticletitle{Fast R-CNN}. In \bibinfo{booktitle}{\emph{Proceedings of the IEEE International Conference on Computer Vision (ICCV)}}.
\newblock


\bibitem[Guo et~al\mbox{.}(2024)]%
        {guoembedding}
\bibfield{author}{\bibinfo{person}{Xingzhuo Guo}, \bibinfo{person}{Junwei Pan}, \bibinfo{person}{Ximei Wang}, \bibinfo{person}{Baixu Chen}, \bibinfo{person}{Jie Jiang}, {and} \bibinfo{person}{Mingsheng Long}.} \bibinfo{year}{2024}\natexlab{}.
\newblock \showarticletitle{On the Embedding Collapse when Scaling up Recommendation Models}. In \bibinfo{booktitle}{\emph{Forty-first International Conference on Machine Learning}}.
\newblock


\bibitem[Jegou et~al\mbox{.}(2010)]%
        {jegou2010product}
\bibfield{author}{\bibinfo{person}{Herve Jegou}, \bibinfo{person}{Matthijs Douze}, {and} \bibinfo{person}{Cordelia Schmid}.} \bibinfo{year}{2010}\natexlab{}.
\newblock \showarticletitle{Product quantization for nearest neighbor search}.
\newblock \bibinfo{journal}{\emph{IEEE Transactions on Pattern Analysis and Machine Intelligence}} \bibinfo{volume}{33}, \bibinfo{number}{1} (\bibinfo{year}{2010}), \bibinfo{pages}{117--128}.
\newblock


\bibitem[Lample et~al\mbox{.}(2019)]%
        {lample2019large}
\bibfield{author}{\bibinfo{person}{Guillaume Lample}, \bibinfo{person}{Alexandre Sablayrolles}, \bibinfo{person}{Marc'Aurelio Ranzato}, \bibinfo{person}{Ludovic Denoyer}, {and} \bibinfo{person}{Herv{\'e} J{\'e}gou}.} \bibinfo{year}{2019}\natexlab{}.
\newblock \showarticletitle{Large memory layers with product keys}.
\newblock \bibinfo{journal}{\emph{Advances in Neural Information Processing Systems}}  \bibinfo{volume}{32} (\bibinfo{year}{2019}).
\newblock


\bibitem[Liu et~al\mbox{.}(2023)]%
        {liu2023deep}
\bibfield{author}{\bibinfo{person}{Qi Liu}, \bibinfo{person}{Xuyang Hou}, \bibinfo{person}{Haoran Jin}, \bibinfo{person}{Zhe Wang}, \bibinfo{person}{Defu Lian}, \bibinfo{person}{Tan Qu}, \bibinfo{person}{Jia Cheng}, \bibinfo{person}{Jun Lei}, {et~al\mbox{.}}} \bibinfo{year}{2023}\natexlab{}.
\newblock \showarticletitle{Deep Group Interest Modeling of Full Lifelong User Behaviors for CTR Prediction}.
\newblock \bibinfo{journal}{\emph{arXiv preprint arXiv:2311.10764}} (\bibinfo{year}{2023}).
\newblock


\bibitem[Miller et~al\mbox{.}(2016)]%
        {miller2016key}
\bibfield{author}{\bibinfo{person}{Alexander Miller}, \bibinfo{person}{Adam Fisch}, \bibinfo{person}{Jesse Dodge}, \bibinfo{person}{Amir-Hossein Karimi}, \bibinfo{person}{Antoine Bordes}, {and} \bibinfo{person}{Jason Weston}.} \bibinfo{year}{2016}\natexlab{}.
\newblock \showarticletitle{Key-Value Memory Networks for Directly Reading Documents}. In \bibinfo{booktitle}{\emph{Proceedings of the 2016 Conference on Empirical Methods in Natural Language Processing}}.
\newblock


\bibitem[Pi et~al\mbox{.}(2019)]%
        {pi2019practice}
\bibfield{author}{\bibinfo{person}{Qi Pi}, \bibinfo{person}{Weijie Bian}, \bibinfo{person}{Guorui Zhou}, \bibinfo{person}{Xiaoqiang Zhu}, {and} \bibinfo{person}{Kun Gai}.} \bibinfo{year}{2019}\natexlab{}.
\newblock \showarticletitle{Practice on long sequential user behavior modeling for click-through rate prediction}. In \bibinfo{booktitle}{\emph{Proceedings of the 25th ACM SIGKDD International Conference on Knowledge Discovery \& Data Mining}}. \bibinfo{pages}{2671--2679}.
\newblock


\bibitem[Pi et~al\mbox{.}(2020)]%
        {pi2020search}
\bibfield{author}{\bibinfo{person}{Qi Pi}, \bibinfo{person}{Guorui Zhou}, \bibinfo{person}{Yujing Zhang}, \bibinfo{person}{Zhe Wang}, \bibinfo{person}{Lejian Ren}, \bibinfo{person}{Ying Fan}, \bibinfo{person}{Xiaoqiang Zhu}, {and} \bibinfo{person}{Kun Gai}.} \bibinfo{year}{2020}\natexlab{}.
\newblock \showarticletitle{Search-based user interest modeling with lifelong sequential behavior data for click-through rate prediction}. In \bibinfo{booktitle}{\emph{Proceedings of the 29th ACM International Conference on Information \& Knowledge Management}}. \bibinfo{pages}{2685--2692}.
\newblock


\bibitem[Yan et~al\mbox{.}(2014)]%
        {yan2014coupled}
\bibfield{author}{\bibinfo{person}{Ling Yan}, \bibinfo{person}{Wu-Jun Li}, \bibinfo{person}{Gui-Rong Xue}, {and} \bibinfo{person}{Dingyi Han}.} \bibinfo{year}{2014}\natexlab{}.
\newblock \showarticletitle{Coupled group lasso for web-scale ctr prediction in display advertising}. In \bibinfo{booktitle}{\emph{International conference on machine learning}}. PMLR, \bibinfo{pages}{802--810}.
\newblock


\bibitem[Yu et~al\mbox{.}(2024)]%
        {yu2024ifa}
\bibfield{author}{\bibinfo{person}{Wenhui Yu}, \bibinfo{person}{Chao Feng}, \bibinfo{person}{Yanze Zhang}, \bibinfo{person}{Lantao Hu}, \bibinfo{person}{Peng Jiang}, {and} \bibinfo{person}{Han Li}.} \bibinfo{year}{2024}\natexlab{}.
\newblock \showarticletitle{IFA: Interaction Fidelity Attention for Entire Lifelong Behaviour Sequence Modeling}.
\newblock \bibinfo{journal}{\emph{arXiv preprint arXiv:2406.09742}} (\bibinfo{year}{2024}).
\newblock


\bibitem[Zhai et~al\mbox{.}({[n.\,d.]})]%
        {zhaiactions}
\bibfield{author}{\bibinfo{person}{Jiaqi Zhai}, \bibinfo{person}{Lucy Liao}, \bibinfo{person}{Xing Liu}, \bibinfo{person}{Yueming Wang}, \bibinfo{person}{Rui Li}, \bibinfo{person}{Xuan Cao}, \bibinfo{person}{Leon Gao}, \bibinfo{person}{Zhaojie Gong}, \bibinfo{person}{Fangda Gu}, \bibinfo{person}{Jiayuan He}, {et~al\mbox{.}}} \bibinfo{year}{[n.\,d.]}\natexlab{}.
\newblock \showarticletitle{Actions Speak Louder than Words: Trillion-Parameter Sequential Transducers for Generative Recommendations}. In \bibinfo{booktitle}{\emph{International conference on machine learning}}.
\newblock


\bibitem[Zhang et~al\mbox{.}({[n.\,d.]})]%
        {zhangwukong}
\bibfield{author}{\bibinfo{person}{Buyun Zhang}, \bibinfo{person}{Liang Luo}, \bibinfo{person}{Yuxin Chen}, \bibinfo{person}{Jade Nie}, \bibinfo{person}{Xi Liu}, \bibinfo{person}{Shen Li}, \bibinfo{person}{Yanli Zhao}, \bibinfo{person}{Yuchen Hao}, \bibinfo{person}{Yantao Yao}, \bibinfo{person}{Ellie~Dingqiao Wen}, {et~al\mbox{.}}} \bibinfo{year}{[n.\,d.]}\natexlab{}.
\newblock \showarticletitle{Wukong: Towards a Scaling Law for Large-Scale Recommendation}. In \bibinfo{booktitle}{\emph{Forty-first International Conference on Machine Learning}}.
\newblock


\bibitem[Zhou et~al\mbox{.}(2019)]%
        {zhou2019deep}
\bibfield{author}{\bibinfo{person}{Guorui Zhou}, \bibinfo{person}{Na Mou}, \bibinfo{person}{Ying Fan}, \bibinfo{person}{Qi Pi}, \bibinfo{person}{Weijie Bian}, \bibinfo{person}{Chang Zhou}, \bibinfo{person}{Xiaoqiang Zhu}, {and} \bibinfo{person}{Kun Gai}.} \bibinfo{year}{2019}\natexlab{}.
\newblock \showarticletitle{Deep interest evolution network for click-through rate prediction}. In \bibinfo{booktitle}{\emph{Proceedings of the AAAI conference on artificial intelligence}}, Vol.~\bibinfo{volume}{33}. \bibinfo{pages}{5941--5948}.
\newblock


\bibitem[Zhou et~al\mbox{.}(2018)]%
        {zhou2018deep}
\bibfield{author}{\bibinfo{person}{Guorui Zhou}, \bibinfo{person}{Xiaoqiang Zhu}, \bibinfo{person}{Chenru Song}, \bibinfo{person}{Ying Fan}, \bibinfo{person}{Han Zhu}, \bibinfo{person}{Xiao Ma}, \bibinfo{person}{Yanghui Yan}, \bibinfo{person}{Junqi Jin}, \bibinfo{person}{Han Li}, {and} \bibinfo{person}{Kun Gai}.} \bibinfo{year}{2018}\natexlab{}.
\newblock \showarticletitle{Deep interest network for click-through rate prediction}. In \bibinfo{booktitle}{\emph{Proceedings of the 24th ACM SIGKDD international conference on knowledge discovery \& data mining}}. \bibinfo{pages}{1059--1068}.
\newblock


\end{thebibliography}

\end{document}